\documentclass[conference]{IEEEtran}
\IEEEoverridecommandlockouts
\usepackage{cite}
\usepackage{amsmath,amssymb,amsfonts}
\usepackage{algorithmic}
\usepackage{graphicx}
\usepackage{tabularx} 
\usepackage{textcomp}
\usepackage{xcolor}
\usepackage{subcaption}
\usepackage{hyperref}
\usepackage{wrapfig}

\def\BibTeX{{\rm B\kern-.05em{\sc i\kern-.025em b}\kern-.08em
T\kern-.1667em\lower.7ex\hbox{E}\kern-.125emX}}
\begin{document}

\title{Bone Fracture Classification Using Transfer Learning}

\author{\IEEEauthorblockN{Shyam Gupta}
\IEEEauthorblockA{\textit{Technicshe Universitat Dortmund} \\
Dortmund, Germany \\
shyam.gupta@tu-dortmund.de}
\and
\IEEEauthorblockN{Dhanisha Sharma}
\IEEEauthorblockA{\textit{B.Sc Honors Physics} \\
\textit{Devi Ahilya Vishwa Vidyalaya }\\
Indore, India \\
dhanisha522292@gmail.com}
}

\maketitle

\begin{abstract}
The manual examination of X-ray images for fractures is a time-consuming process that is prone to human error. In this work, we introduce a robust yet simple training loop for the classification of fractures, which significantly outperforms existing methods. Our method achieves superior performance in less than ten epochs and utilizes the latest dataset to deliver the best-performing model for this task. We emphasize the importance of training deep learning models responsibly and efficiently, as well as the critical role of selecting high-quality datasets.

\end{abstract}


\section{Introduction}
Bone fractures are a common medical condition that significantly impact patients' health and quality of life. Accurate and timely diagnosis of fractures is crucial for effective treatment and recovery. Traditionally, radiologists and orthopedic surgeons have relied on X-ray imaging to detect fractures. However, interpreting these images can be challenging, particularly when dealing with minute or complex fractures that are not easily visible. Misdiagnosis or delayed diagnosis can lead to improper treatment, prolonged recovery times, and increased healthcare costs.

\section{Problem Statement}
The problem is exacerbated in busy clinical settings, where radiologists may face a high volume of images to review. Additionally, subtle fractures may be overlooked due to the limitations of human perception. Therefore, there is a critical need for automated methods that can assist healthcare professionals in accurately identifying fractures in X-ray images, thereby improving diagnostic accuracy and efficiency.

\section{Related Works}

The study by \cite{yadav2020bone} presents a deep neural network (DNN) model for classifying fractured and healthy bones, addressing the limitations of manual X-ray diagnosis, which is time-consuming and error-prone. To counteract overfitting on small datasets, data augmentation techniques were employed, significantly enhancing model performance. The model achieved a classification accuracy of 92.44\% using 5-fold cross-validation, surpassing previous methods with accuracies of 84.7\% and 86\%. This advancement underscores the potential of automated DNN-based systems in improving diagnostic accuracy and efficiency in bone fracture detection.

In recent research, Yadav et al. (2022) proposed a novel hybrid model, the hybrid scale fracture network (SFNet), which integrates convolutional neural networks (CNN) with an improved canny edge algorithm for bone fracture diagnosis \cite{yadav2022hybrid}. This model efficiently identifies fractures by using multi-scale feature fusion, achieving impressive training accuracy of 99.12\% which was very close to our results, respectively. The SFNet model's innovation lies in its combination of grey images and canny edge images for training, demonstrating superior performance compared to other deep CNN models, thus advancing the capabilities in medical image diagnosis.

The development of automated fracture detection systems is crucial for enhancing computer-aided telemedicine. Ma and Luo (2020) introduce Crack-Sensitive Convolutional Neural Network (CrackNet) within a two-stage system for fracture detection, demonstrating superior performance. Initially, Faster R-CNN identifies 20 bone regions in X-ray images, followed by CrackNet assessing each region for fractures. Tested on 1052 images, including 526 fractured cases, their system achieved 90.11\% accuracy and 90.14\% F-measure, outperforming other two-stage systems. This approach significantly aids hospitals lacking experienced surgeons by providing accurate and efficient fracture diagnosis \cite{ma2020bone}.

\section{Ease of Use}

\subsection{Enhanced Diagnostic Accuracy:}
The accuracy of human radiograph experts in classifying bone fractures varies across studies, but it typically ranges from 77.5\% to 93.5\%. Our proposed algorithm significantly improves the accuracy of fracture detection, reducing the likelihood of missed fractures, particularly those that are minute or complex. By leveraging EfficientNet's sophisticated architecture, the model can identify subtle patterns and features in X-ray images that may be challenging for the human eye to detect.

\subsection{Time Efficiency:}

Automated fracture classification allows for quicker analysis of X-ray images, enabling doctors to focus on diagnosis and treatment rather than manual image review. In emergency settings, rapid identification of fractures can be crucial for timely medical intervention.

\subsection{Consistency and Reliability:}
The algorithm provides consistent results, minimizing the variability inherent in human interpretation of X-ray images. This consistency ensures that every patient receives the same high standard of diagnostic care.

\section{Evaluation Metrics}

\subsection{Precision}
\begin{equation}
\text{Precision} = \frac{TP}{TP + FP}
\end{equation}

\subsection{Recall}
\begin{equation}
\text{Recall} = \frac{TP}{TP + FN}
\end{equation}

\subsection{F1 Score}
\begin{equation}
F1 = 2 \cdot \frac{\text{Precision} \cdot \text{Recall}}{\text{Precision} + \text{Recall}}
\end{equation}

\subsection{AUC ROC}
The AUC (Area Under the Curve) ROC (Receiver Operating Characteristic) is computed from the ROC curve, which plots the true positive rate (TPR) against the false positive rate (FPR) at various threshold settings.

\subsection{Accuracy}
\begin{equation}
\text{Accuracy} = \frac{TP + TN}{TP + TN + FP + FN}
\end{equation}\newline

\section{Objective \& Methodology}
This paper aims to address the challenge of bone fracture classification by leveraging deep learning techniques, specifically utilizing EfficientNet for transfer learning. EfficientNet, known for its balance between accuracy and computational efficiency, presents an ideal candidate for developing a robust fracture detection model. The goal is to create a system that can automatically classify X-ray images to determine the presence or absence of fractures with high accuracy, aiding doctors in making quicker and more accurate diagnoses.

\subsection{Dataset - Fracatlas}
The Frac-Atlas dataset is an extensive and meticulously curated collection of X-ray images specifically designed for the detection of bone fractures. Unlike other existing fracture datasets, Frac-Atlas encompasses a wide variety of fracture types, including both common and rare fractures, as well as a range of complexities from clear, easily visible fractures to subtle, minute ones. This diversity ensures that the dataset provides a comprehensive training ground for developing robust fracture detection algorithms. Furthermore, Frac-Atlas includes high-resolution images with annotated fracture locations, making it an invaluable resource for training deep learning models. The dataset also includes metadata such as patient demographics, fracture types, and severity, which can be used for more detailed analysis and model training. This level of detail and variety makes Frac-Atlas a superior dataset for advancing the state of fracture detection technology, providing a rich foundation for both academic research and practical application in clinical settings.\cite{FracatlasPWC}

Below are some techniques we tried to use on our dataset to solve and explore the problem, we used such techniques which can help model learn differentiation between classes, based on contract, edges, color diffusion, size scaling,etc\ldots.

\subsection{CLAHE Transform (Contrast Limited Adaptive Histogram Equalization)}
CLAHE enhances X-ray image contrast by applying adaptive histogram equalization on small regions, preventing noise amplification. This technique improves the visibility of subtle features like hairline fractures by redistributing pixel intensity values, making fractures more prominent and easier to detect for both human eyes and algorithms.

\begin{figure}[htbp]
    \centering
    \begin{subfigure}[b]{0.45\textwidth}
        \includegraphics[width=\textwidth]{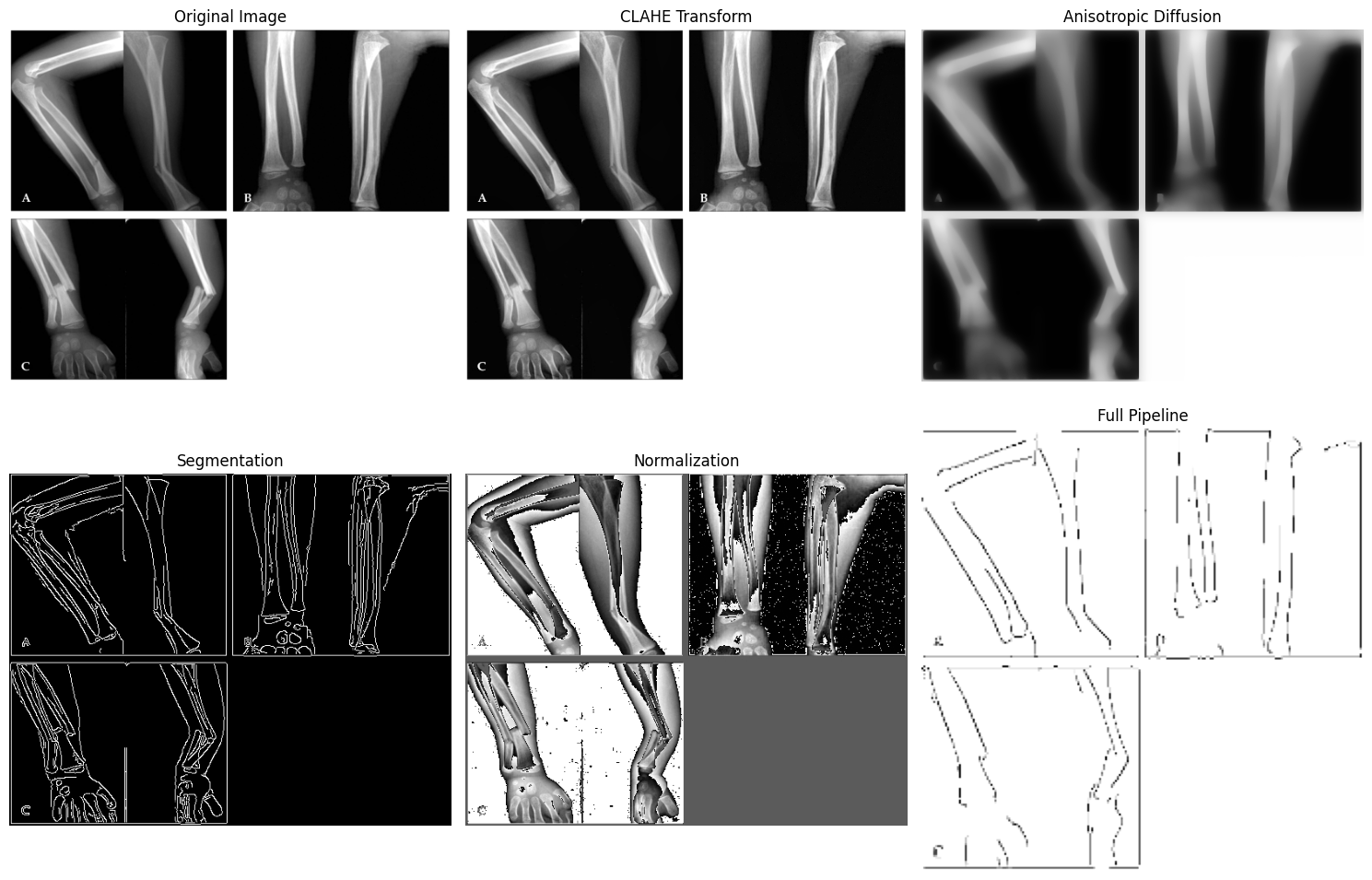}
    \end{subfigure}
    \hfill
    \begin{subfigure}[b]{0.45\textwidth}
        \includegraphics[width=\textwidth]{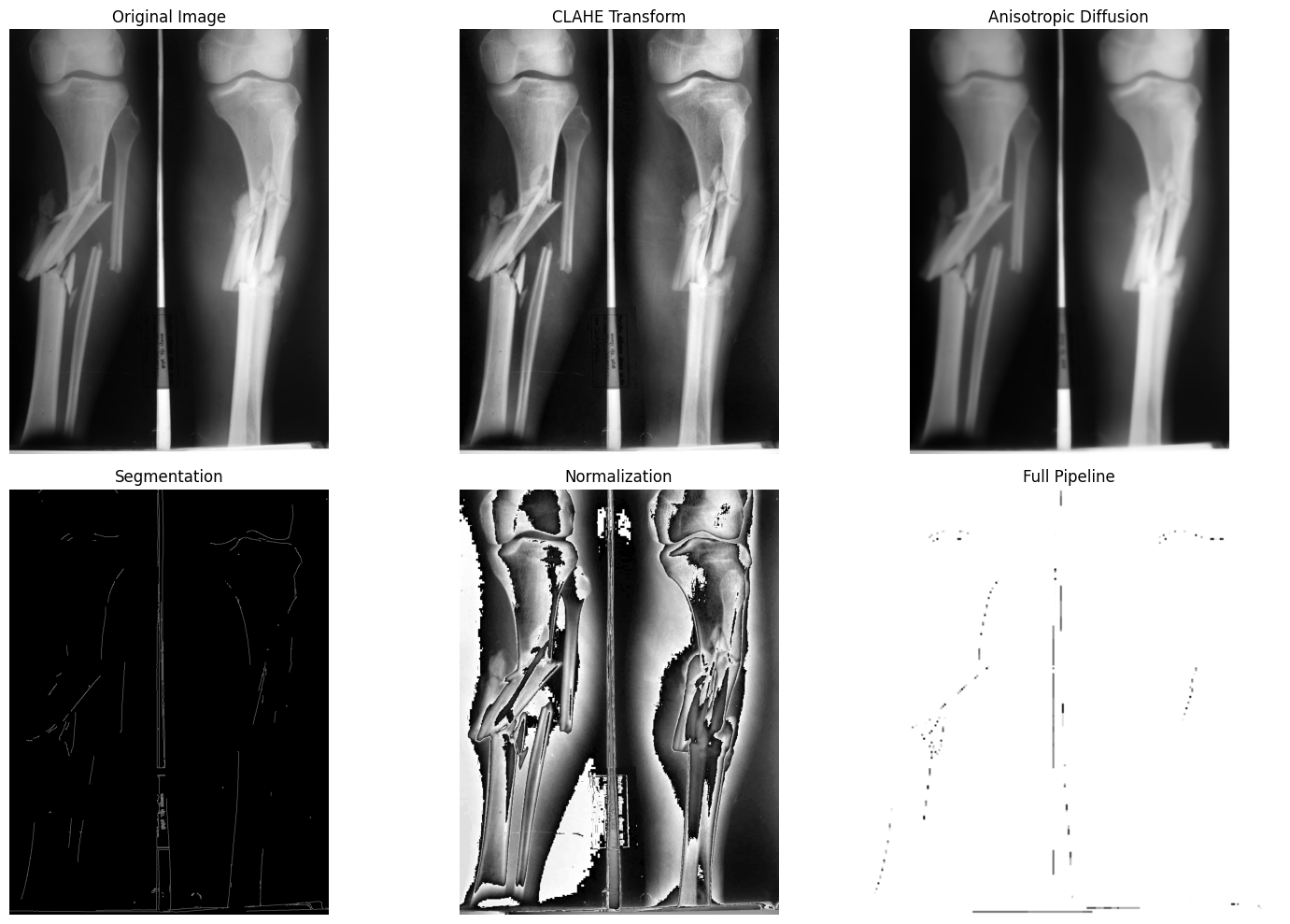}
        \caption{Ineffective Transformations}
\end{subfigure}
    \caption{We identified using transformations did not help us identify fractures neither helped us to get good results. Thus, making our training loop simpler and easier to learn/generalize on raw xray-images}
    \label{fig:sidebyside}
\end{figure}

\subsection{Anisotropic Diffusion Transform}
Anisotropic Diffusion reduces image noise while preserving important edges. By controlling diffusion to smooth homogeneous regions more than edges, this technique enhances fracture clarity without blurring, making fractures more distinguishable and maintaining diagnostic quality.

\subsection{Canny Edge Detection}
Canny Edge Detection identifies boundaries within X-ray images by detecting sharp intensity changes. It involves noise reduction, gradient calculation, non-maximum suppression, and edge tracking to highlight bone structures and potential fracture lines, making it easier to identify fractures in complex images.

\subsection{Image Segmentation Transform}
Image Segmentation divides an image into multiple segments, isolating regions of interest like specific bones or suspected fracture areas. This focused analysis reduces complexity and improves fracture detection accuracy, highlighting high-probability fracture areas and enhancing overall model performance.

\subsection{Random Central Cropping}
Random Central Cropping is a data augmentation technique that creates multiple variations of an X-ray image by cropping the central region. This improves model robustness by exposing it to different views, preventing overfitting, and enhancing generalization to new images, especially useful when labeled data is limited.

\section{Training \& Results}

Despite our thorough experimentation with various preprocessing techniques, including CLAHE Transform, Anisotropic Diffusion Transform, Canny Edge Detection, Image Segmentation Transform, and Random Central Cropping, these methods did not significantly enhance the model's ability to detect fractures. While each technique offered potential advantages for highlighting fractures, they also introduced complexity and noise that ultimately did not improve detection accuracy.

\begin{figure}[htbp]
    \centering
    \begin{subfigure}[b]{0.45\textwidth}
        \includegraphics[width=\textwidth]{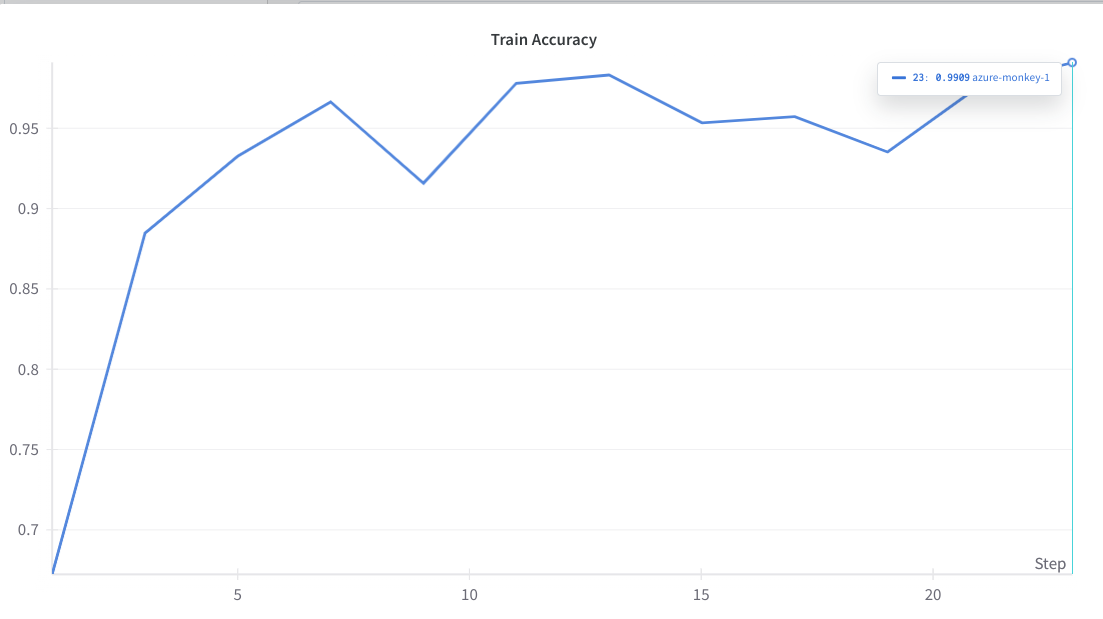}
        \caption{Training Curve}
        \label{fig:fractured}
    \end{subfigure}
    \hfill
    \begin{subfigure}[b]{0.45\textwidth}
        \includegraphics[width=\textwidth]{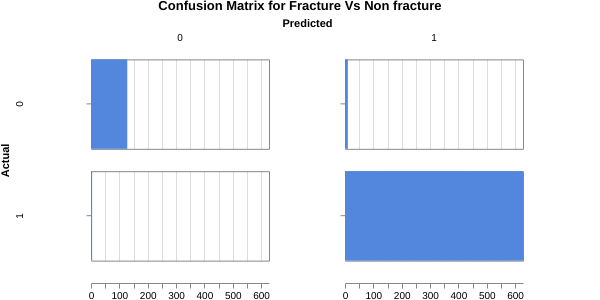}
        \caption{Confusion Matrix}
        \label{fig:nonfractured}
    \end{subfigure}
    \caption{Metrics}
    \label{fig:sidebyside}
\end{figure}

\begin{figure}[htbp]
    \centering
    \begin{subfigure}[b]{0.45\textwidth}
        \includegraphics[width=\textwidth]{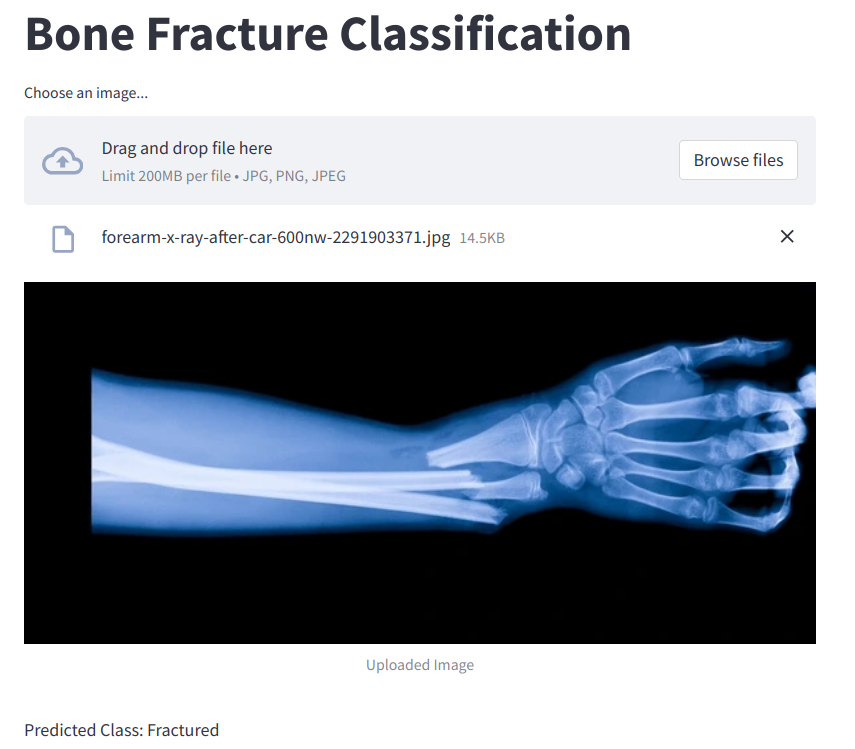}
        \caption{Classification of Fractured Bones}
        \label{fig:fractured}
    \end{subfigure}
    \hfill
    \begin{subfigure}[b]{0.45\textwidth}
        \includegraphics[width=\textwidth]{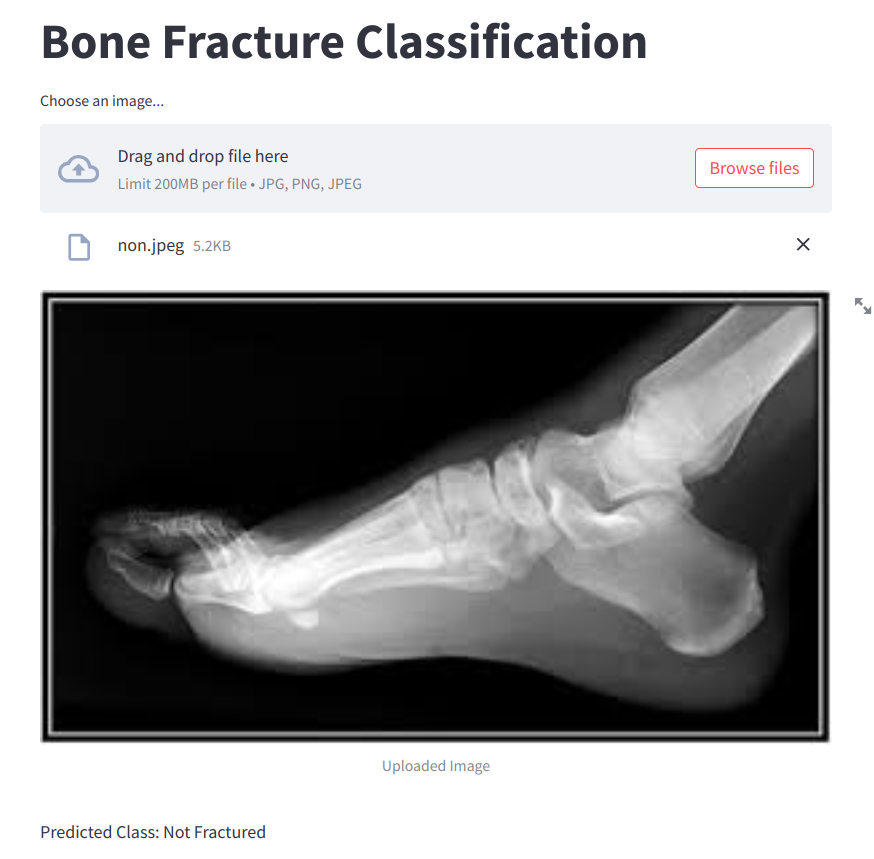}
        \caption{Classification of Non-Fractured Bones}
        \label{fig:nonfractured}
    \end{subfigure}
    \caption{Results From Deployed Model}
    \label{fig:sidebyside}
\end{figure}

Instead, we found that normalizing the X-ray images alone provided the best results. Normalization adjusted the pixel intensity values to a consistent scale, simplifying the image data and enhancing the model's ability to learn relevant features without additional noise or artifacts. This streamlined approach proved to be highly effective.

No \textit{center cropping}! while cropping the images the fractures located at the corners went out of the convolution window. As for computation, we used \textsl{Tesla-P100 GPU} for training, freely available on Kaggle. Our model, built using EfficientNet-B6 with pretrained weights, was trained with a batch size of 16, with \textsl{CrossEntropyLoss}, optimizer as \textsl{Adam} and scheduler as \textsl{Cosine Annealing Warm Restart optimizer}. 

A point to note is, while we trained on larger batch sizes, the accuracy was increasing slowly, so we figured that larger batch sizes lead to slower training improvements.  This setup facilitated efficient learning and convergence. The training process was remarkably quick, achieving an impressive \textit{training accuracy of 99\%, with validation accuracy 97\%}  on the training set in just \textbf{7 epochs}. The model also demonstrated robust performance on the testing set, achieving an \textbf{Testing accuracy of 96.83\%}.

\begin{table}[htbp]
\caption{Comparison of Bone Fracture Classification Models}
\begin{center}
\begin{tabular}{|c|c|c|}
\hline
\textbf{Authors} & \textbf{Publication Year} & \textbf{Testing Accuracy (\%)} \\ \hline
Muhammet Emin Sahin & 2023  & 86 \\ \hline
Pongsakorn Samothai et al. & 2022 & 88 \\ \hline
D. P. Yadav et al. & 2020 & 95 \\ \hline
Basha et al. & 2020 & 92 \\ \hline
Rao, L.J. et al. & 2020 & 90 \\ \hline
\textbf{Our method} & \textbf{2024} & \textbf{96.83} \\ \hline
\end{tabular}
\label{tab1}
\end{center}
\end{table}

As for some post training validation figures, we achieved \textbf{Test Precision: 0.9770, Test Recall: 0.9606, Test F1 Score: 0.9686, Test AUC ROC: 0.9606}. Which completely outperforms, existing methods of fracture detection, and provides a more robust solution.

These results underscore the effectiveness of our approach, highlighting the power of EfficientNet-B6 and the importance of choosing appropriate preprocessing techniques. The high accuracy achieved in both training and testing phases indicates the model's robustness and potential for practical application in clinical settings, significantly aiding in the accurate and efficient diagnosis of bone fractures.

\section{Conclusion and Future scopes}
We developed a simple yet effective method for detecting fractures by leveraging transfer learning and the power of data augmentation. Our approach demonstrates the high quality of the ready-to-train images available in the \textit{FracAtlas} dataset. By emphasizing the initial step of the data science pipeline, namely \textit{data}, we achieved remarkable results. The simplicity of our model, combined with the superior quality of the dataset, enabled us to achieve over 99\% training accuracy in just seven epochs and a testing accuracy of 96.83

The \textit{FracAtlas} dataset also provides bounding box coordinates to detect the exact location of fractures. We propose using the same network as a backbone to train a region localization algorithm for precise fracture detection. Addressing the imbalance in the dataset remains a challenge that we must tackle in future work.

\bibliographystyle{plain}
\bibliography{references}

\begin{thebibliography}{99}

\bibitem{meena2022radio}
Meena, T., \& Roy, S. (2022). Bone Fracture Detection Using Deep Supervised Learning from Radiological Images: A Paradigm Shift. Diagnostics, 12 \href{https://doi.org/10.3390/diagnostics12102420}{\color{blue}doi-link}. 

\bibitem{yadav2022hybrid}
Yadav, D.P., Sharma, A., Athithan, S., Bhola, A., Sharma, B., \& Dhaou, I.B. (2022). Hybrid SFNet Model for Bone Fracture Detection and Classification Using ML/DL. \textit{Sensors}, 22. \href{https://doi.org/10.3390/s22155823}{\color{blue}doi-link}.

\bibitem{samothai2022evaluation}
Samothai, P., Sanguansat, P., Kheaksong, A., Srisomboon, K., \& Lee, W. (2022). The Evaluation of Bone Fracture Detection of YOLO Series. In \textit{37th International Technical Conference on Circuits/Systems, Computers and Communications (ITC-CSCC)}, pp. 1054-1057. \href{https://doi.org/10.1109/ITC-CSCC55581.2022.9895016}{\color{blue}doi-link}.

\bibitem{santhiya2022bone}
Santhiya, J., \& Ebenezer, M.P. (2022). Bone Fracture Detection Using Python. [Online]. Available: \href{https://doi.org/10.3390/s22155823}{\color{blue}doi-link}.

\bibitem{santos2022feasibility}
Santos, K.C., Fernandes, C.A., \& Costa, J.R. (2022). Feasibility of Bone Fracture Detection Using Microwave Imaging. \textit{IEEE Open Journal of Antennas and Propagation}, 3, 836-847. \href{https://doi.org/10.1109/OJAP.2022.3194217}{\color{blue}doi-link}.

\bibitem{alghaithi2021artificial}
AlGhaithi, A., \& al Maskari, S. (2021). Artificial intelligence application in bone fracture detection. \textit{Journal of Musculoskeletal Surgery and Research}, 5, 4-9. \href{https://doi.org/10.4103/jmsr.jmsr_132_20}{\color{blue}doi-link}.

\bibitem{yadav2020bone}
Yadav, D.P., \& Rathor, S. (2020). Bone Fracture Detection and Classification using Deep Learning Approach. In \textit{International Conference on Power Electronics \& IoT Applications in Renewable Energy and its Control (PARC)}, pp. 282-285. \href{https://doi.org/10.1109/PARC49193.2020.236611}{\color{blue}doi-link}.

\bibitem{basha2020enhanced}
Basha, C.Z., Reddy, M., Nikhil, K.H., Venkatesh, P.M., \& Asish, A.V. (2020). Enhanced Computer Aided Bone Fracture Detection Employing X-ray Images by Harris Corner Technique. In \textit{Fourth International Conference on Computing Methodologies and Communication (ICCMC)}, pp. 991-995. \href{https://doi: 10.1109/ICCMC48092.2020.ICCMC-000184}{\color{blue}doi-link}.

\bibitem{sahin2023image}
Sahin, M.E. (2023). Image processing and machine learning-based bone fracture detection and classification using X-ray images. \textit{International Journal of Imaging Systems and Technology}, 33, 853-865. \href{https://doi.org/10.1002/ima.22849}{\color{blue}doi-link}.

\bibitem{ma2020bone}
Ma, Y., \& Luo, Y. (2020). Bone fracture detection through the two-stage system of Crack-Sensitive Convolutional Neural Network. \textit{Informatics in Medicine Unlocked}, 100452. \href{https://doi.org/10.1016/j.imu.2020.100452}{\color{blue}doi-link}.

\bibitem{zhang2020new}
Zhang, X., Wang, Y., Cheng, C., Lu, L., Xiao, J., Liao, C., \& Miao, S. (2020). A New Window Loss Function for Bone Fracture Detection and Localization in X-ray Images with Point-based Annotation. \href{https://doi.org/10.48550/arXiv.2012.04066}{\color{blue}doi-link}.

\bibitem{rao2020effective}
Rao, L.J., Neelakanteswar, P., Ramkumar, M., Krishna, A., \& Basha, C.Z. (2020). An Effective Bone Fracture Detection using Bag-of-Visual-Words with the Features Extracted from SIFT. In \textit{International Conference on Electronics and Sustainable Communication Systems (ICESC)}, pp. 6-10. \href{https://doi.org/10.1109/ICESC48915.2020.9156035}{\color{blue}doi-link}.

\bibitem{yang2019long}
Yang, A.Y., \& Cheng, L. (2019). Long-Bone Fracture Detection using Artificial Neural Networks based on Line Features of X-ray Images. In \textit{IEEE Symposium Series on Computational Intelligence (SSCI)}, pp. 2595-2602. \href{https://doi.org/10.1109/SSCI44817.2019.9002664}{\color{blue}doi-link}.

\bibitem{basha2019effective}
Basha, C.Z., Padmaja, T.M., \& Balaji, G.N. (2019). An Effective and Reliable Computer Automated Technique for Bone Fracture Detection. \textit{EAI Endorsed Transactions on Pervasive Health and Technology}, 5(18). \href{https://doi.org/10.4108/eai.13-7-2018.162402}{\color{blue}doi-link}.

\bibitem{FracatlasPWC}
Fracatlas dataset \href{https://paperswithcode.com/dataset/fracatlas}{\color{blue}dataset}.

\bibitem{kaggle}
Kaggle \href{https://kaggle.com}{\color{blue}link}


\end{thebibliography}

\end{document}